\documentclass[]{aastex631}

\begin{document}

\title{Mitigating Cosmic Microwave Background Shadow Degradation of Tensor-to-scalar Ratio Measurements through Map-based Studies 
}

\author{Tamaki Murokoshi}
\affiliation{Astronomical Institute, Tohoku University, Sendai, Miyagi 980-8578, Japan}
\author{Yuji Chinone}
\affiliation{QUP (WPI), KEK, Oho 1-1, Tsukuba, Ibaraki 305-0801, Japan}
\affiliation{Kavli Institute for the Physics and Mathematics of the Universe (WPI), UTIAS, The University of Tokyo, Kashiwa, Chiba, 277-8583, Japan}
\author{Masashi Nashimoto}
\affiliation{Department of Astronomy, Graduate School of Science, The University of Tokyo, 7-3-1 Hongo, Bunkyo-ku, Tokyo 113-0033, Japan}
\author{Kiyotomo Ichiki}
\affiliation{Graduate School of Science, Division of Particle and Astrophysical Science, Nagoya University, Furocho, Chikusa-ku, Nagoya, Aichi 464-8602, Japan}
\affiliation{Kobayashi-Maskawa Institute for the Origin of Particles and the Universe, Nagoya University, Furocho, Chikusa-ku, Nagoya, Aichi 464-8602, Japan}
\affiliation{Institute for Advanced Research, Nagoya University, Furocho, Chikusa-ku, Nagoya, Aichi 464-8602, Japan}
\author{Makoto Hattori}
\affiliation{Astronomical Institute, Tohoku University, Sendai, Miyagi 980-8578, Japan}

\begin{abstract}
It has been pointed out that the spurious cosmic microwave background (CMB) $B$-mode polarization signals caused by the absorption of the CMB monopole component due to the Galactic interstellar matter, called the CMB shadow, degrade the accuracy of detecting the CMB $B$-mode polarization signals imprinted by primordial gravitational waves.
We have made a realistic estimation using simulated sky maps of how the CMB shadow affects forthcoming high-precision CMB $B$-mode experiments for the first time.
The Delta-map method, an internal template method taking into account the first-order spatial variation of foregrounds' spectral parameters, is applied as a foreground removal method.
We show that  if the CMB shadow effects are not taken into account in the foreground removal process, future observations would lead  to the  false detection of  the CMB $B$-mode polarization signals originating from primordial gravitational waves.
We also show that the effect of the CMB shadow can be mitigated by our revised Delta-map method to target the CMB $B$-mode polarization signals at the level of tensor-to-scalar ratio $r=0.001$.
\end{abstract}

\section{Introduction} \label{sec:intro}
One of the current hot topics of modern observational cosmology is observational confirmation 
of the inflation theory \citep{Guth1989, Sato1981} by detecting 
the cosmic microwave background (CMB) $B$-mode polarization signals (e.g., Simons Array, \citealt{SA}; Simons Observatory, \citealt{Ade_2019}; LiteBIRD, \citealt{Hazumi_2020}; BICEP3, \citealt{BICEP3}; CMB-S4, \citealt{snowmass_s4}) generated by primordial gravitational waves 
\citep{Zaldarriaga_1997,Seljak_1997}. 
The signals have not been detected yet. 
The current upper limit of tensor-to-scalar ratio $r$ is $r < 0.032$ at 95\% CL \citep{Tristram2022}.
The next generations of the CMB polarization experiments target the sensitivity to $r\sim 0.001$ (e.g., Simons Observatory, \citealt{Ade_2019}; LiteBIRD, \citealt{Hazumi_2020}; BICEP3, \citealt{BICEP3}; CMB-S4, \citealt{snowmass_s4}). 
Polarized emissions originating from celestial objects, especially the Galactic synchrotron emission and thermal emission from the Galactic interstellar dust, are the main obstacles to realizing such highly sensitive measurements \citep[e.g.,][]{Ichiki_2014}. 
Many efforts have been paid to improve the foreground removal accuracy 
\citep[e.g.,][]{Delabrouille_2003,Eriksen2008, Delabrouille_2008, Remazeilles_2011, Fern_ndez_Cobos_2012}. 
\citet{Nashimoto2020} pointed out that the 
spurious CMB polarization signals generated by absorption of the CMB monopole component by the interstellar medium, called the CMB shadow, has a non-negligible effect on detecting the CMB $B$-mode polarization signals at the level of $r\sim 0.001$. 
Their analysis was limited to an order of magnitude estimation using typical values of parameters describing the CMB and foreground components. 
Therefore, a quantitative evaluation of how the CMB shadow degrades the foreground removal accuracy based on realistic, map-based analysis is mandatory.
Furthermore, proposing a method to mitigate the CMB shadow effect and to realize the detection of the CMB $B$-mode polarization signals down to the sensitivity level of $r=0.001$ is required. 

We adopt the Delta-map method \citep{Ichiki2019, Minami+22} for foreground removal.
The Delta-map method is one of the ``internal template'' methods that uses low- and high-frequency bands of the CMB observation maps as ``templates'' for synchrotron emission and thermal dust emission, respectively.
This  method is a moment expansion of the $Q$ and $U$ intensity spectra \citep{Chluba2017} to the first order in the spectral parameters (see Appendix \ref{sec:delta}).
\citet{Ichiki2019} showed that the Delta-map method can reach $r\sim 4\times 10^{-4}$ using noiseless simulated masked sky maps, which consist of polarized CMB, synchrotron emission, and thermal dust emission.
In this paper, firstly, we evaluate how the CMB shadow degrades the accuracy of the tensor-to-scalar ratio $r$ measurements with map-based analysis. 
Next, we implement the CMB shadow in the Delta-map method 
to mitigate it. 

The rest of this paper is organized as follows.
We review the CMB shadow in Sect.\ref{sec:shadow}.
We explain the method of our analysis in Sect.\ref{sec:method}; making simulated sky maps is explained in Sect.\ref{subsec:setup}, and how to mitigate the CMB shadow in the Delta-map method is explained in Sect.\ref{subsec:delta_shadow}.
We show the results of estimating tensor-to-scalar ratio $r$ in Sect.\ref{sec:results}.
Sect.\ref{sec:conclusion} is devoted to discussion.

\section{The CMB shadow} \label{sec:shadow}
The CMB shadow, absorption of the CMB monopole component by the interstellar medium, generates spurious CMB polarization signals.
In this paper, the Stokes $Q$ intensity is defined by the difference in the intensity of the light polarized along the $x$-axis and polarized along the $y$-axis where the direction of the light propagation is taken to the $z$-axis.  
The observed Stokes $Q$ intensity of the polarized CMB light at the frequency $\nu$ propagating through the interstellar medium is given by the solution of the radiative transfer equation \citep[e.g.,][]{radipro} in the optically thin limit as
\begin{eqnarray}
   Q_{\nu}
    &=& Q^{\mathrm{CMB}}_{\nu}
    - \int ds (\alpha_{\nu}^x(s) - \alpha_{\nu}^y(s)) B_{\nu}(T_{\mathrm{CMB}})/2
    + \int ds (j_{\nu}^x(s) - j_{\nu}^y(s)), \label{RadTrans_q}
\end{eqnarray}
where  $Q^{\mathrm{CMB}}_{\nu}$ is the Stokes $Q$ intensity of the polarized CMB signals, $s$ is path length along the ray, $B_{\nu}(T)$ is the Planck function with temperature $T$ at the frequency $\nu$, $T_{\mathrm{CMB}}$ is the CMB temperature ($T_{\mathrm{CMB}}=2.725 \mathrm{K}$ \citealt{Mather_1999}), $j_{\nu}^x(s)$ and $j_{\nu}^y(s)$ are the emissivity of the foreground components, 
and $\alpha_{\nu}^{x}$ and $\alpha_{\nu}^{y}$ are the absorption coefficient of the interstellar matter for the polarization components along the $x$- and the $y$-axes, respectively.
In Equation (\ref{RadTrans_q}), the solution is evaluated up to
the first order of the optical depth $\tau_{\nu}^x=\int ds \:\alpha_{\nu}^x(s)$ and $\tau_{\nu}^y=\int ds \:\alpha_{\nu}^y(s)$. 

The main obstacles of the CMB polarization observations are the Galactic synchrotron emission and the  thermal emission from the Galactic interstellar dust. 
As shown by \citet{Nashimoto2020}, the CMB shadow due to the interstellar dust affects the detection of the CMB $B$-mode polarization signals imprinted by primordial gravitational waves with $r\sim 0.001$.
They also showed that the CMB shadows due to other components, such as the relativistic electrons,
which emit the Galactic synchrotron emission, have negligible effects as far as we target 
$r \geq 0.001$. 
Although spurious polarization signals caused by the absorption of ambient stellar light, in other words, interstellar radiation field, by dust grains also contaminate observed data of the CMB polarization, 
the effect is negligible because the amplitude of the interstellar radiation field is many orders of magnitude smaller than the CMB monopole \citep{Tielens}.
In the following analysis, we only take into account the CMB shadow due to the Galactic interstellar dust and neglect the CMB shadow caused by other components. 
Under this approximation, Equation (\ref{RadTrans_q}) is  
rewritten as   
 \begin{eqnarray}
   Q_{\nu}
    &=& Q^{\mathrm{CMB}}_{\nu}
    - \int ds (\alpha_{\nu}^{\mathrm{d},x}(s) - \alpha_{\nu}^{\mathrm{d},y}(s)) B_{\nu}(T_{\mathrm{CMB}})/2
    + \int ds (\alpha_{\nu}^{\mathrm{d},x}(s) - \alpha_{\nu}^{\mathrm{d},y}(s))B_{\nu}(T_{\mathrm{d}}(s))/2+ Q^{\mathrm{s}}_{\nu}, \label{RadTrans_q2}
\end{eqnarray}
where the emission from the Galactic interstellar dust is assumed as thermal emission with a single temperature $T_{\mathrm{d}}(s)$, $\alpha_{\nu}^{\mathrm{d},x}$ and $\alpha_{\nu}^{\mathrm{d},y}$ are the absorption coefficient of the interstellar dust for the polarization components along the $x$- and the $y$-axes, and $Q^{\mathrm{s}}_{\nu}$ is the Stokes $Q$ intensity of the Galactic synchrotron emission.  
In this study, the attenuation factor of $[1-\exp(-h\nu/k_\mathrm{B}T_\mathrm{d})]$ due to the stimulated emission of dust is not taken into account in the absorption coefficient where $h$ is the Planck constant and $k_\mathrm{B}$ is the Boltzmann constant. 
The reasons why this assumption is safely adopted are explained in Appendix \ref{sec:stiemi}. 
However, it seems that there is yet no consensus on the stimulated emission of dust.
The detection of the CMB shadow and the measurements of its amplitude provide a unique opportunity for  giving the  conclusion in  the debates on the stimulated emission 
of dust grains \citep{Harwit,Bond,Wright}.
We introduce the mean optical depth of dust $\tau^{\mathrm{d}}_{\nu} = (\tau^{\mathrm{d},x}_{\nu} + \tau^{\mathrm{d},y}_{\nu})/2$ and
 the Stokes $Q$ polarization fraction of the thermal dust emission $\Pi_{\nu}^{\mathrm{d},Q}$ as  $\Pi_{\nu}^{\mathrm{d},Q} \tau^{\mathrm{d}}_{\nu} = (\tau_{\nu}^{\mathrm{d},x} - \tau_{\nu}^{\mathrm{d},y})/2$, 
where the optical depth of dust for the polarization components along the $x$- and the $y$-axes are $\tau_{\nu}^{\mathrm{d},x}=\int ds \:\alpha_{\nu}^{\mathrm{d},x}(s)$ and $\tau_{\nu}^{\mathrm{d},y}=\int ds \:\alpha_{\nu}^{\mathrm{d},y}(s)$, respectively.
Then, Equation (\ref{RadTrans_q2}) is reduced to
\begin{eqnarray}
    Q_{\nu}
    &=& Q_{\nu}^{\mathrm{CMB}} 
    - \Pi_{\nu}^{\mathrm{d},Q} \tau^{\mathrm{d}}_{\nu} B_{\nu}(T_{\mathrm{CMB}})
    + \Pi_{\nu}^{\mathrm{d},Q} \tau^{\mathrm{d}}_{\nu} B_{\nu}(T_{\mathrm{d}})+ Q^{\mathrm{s}}_{\nu},
    \label{RadTrans_pol_q} 
\end{eqnarray}
under the assumption that the dust temperature is constant $T_{\mathrm{d}}(s)=T_{\mathrm{d}}$ along each line-of-sight.
Similarly, the Stokes $U$ intensity  is given by 
\begin{eqnarray}
    U_{\nu}
    &=& U_{\nu}^{\mathrm{CMB}}
    - \Pi_{\nu}^{\mathrm{d},U} \tau^{\mathrm{d}}_{\nu} B_{\nu}(T_{\mathrm{CMB}})
    + \Pi_{\nu}^{\mathrm{d},U} \tau^{\mathrm{d}}_{\nu} B_{\nu}(T_{\mathrm{d}})+U^{\mathrm{s}}_{\nu}, \label{RadTrans_pol_u}    
\end{eqnarray}
where $U_{\nu}^{\mathrm{CMB}}$ and $U^{\mathrm{s}}_{\nu}$ are the Stokes $U$ intensity of the polarized CMB signals  and the Galactic synchrotron emission, respectively, and $\Pi_{\nu}^{\mathrm{d},U}$ is the Stokes $U$ polarization fraction of the thermal dust emission. 
In this paper, we model the polarized foreground emission and the CMB shadow according to Eqs.(\ref{RadTrans_pol_q}) and (\ref{RadTrans_pol_u}).

\section{Method} \label{sec:method}

\subsection{Setup of simulated maps} \label{subsec:setup}


Our mock sky maps consist of polarized CMB originating from scalar- and tensor-mode perturbations and the CMB lensing effects, synchrotron emission, thermal dust emission, and the CMB shadow due to dust.
Since one of our purposes is to clarify how the CMB shadow degrades the 
accuracy of the CMB $B$-mode measurements, 
noise is neglected in this analysis.
How the inclusion of the noise degrades the foreground removal accuracy has been studied by \citet{Ichiki2019}. 
The frequency bands are 40, 60, 140, 230, 280, and 340 GHz.
The resolution of the maps is $N_{\mathrm{side}}=4$ in HEALPix \citep{Gorski_2005}.
So we use the reionization bump \citep{Dodelson2}. 
We made 1000 randomly realized maps for the CMB with input tensor-to-scalar ratio $r_{\mathrm{inp}}=0$, 0.001, 0.01, 0.03, 0.1, and 1, respectively, using the Planck 2015 cosmological parameters for ``TT+LowP+lensing'' \citep{PlanckXIII2015}: $\Omega_b h^2=0.02226$, $\Omega_c h^2=0.1186$, $h=0.6781$, $\tau=0.066$, $A_{\mathrm{s}}=2.137\times 10^{-9}$, and $n_{\mathrm{s}}=0.9677$.
We made foreground maps by using the PySM package \citep{Thorne_2017}.
For the synchrotron emission maps, we adopt a single power-law model, ``s1'' in the PySM package, which uses WMAP 9 yr 23 GHz $Q$ and $U$ maps \citep{Bennett_2013} as the reference. 
For the thermal dust emission maps, we adopt a one-component modified blackbody model, ``d1'' in the PySM package, which uses the $Q$ and $U$ maps at 353 GHz in brightness temperature unit obtained from Planck 2015 analysis \citep{PlanckX2015} using the Commander code \citep{Eriksen2004,Eriksen2006,Eriksen2008} as the reference.
The $Q$ and $U$ maps of the CMB shadow due to dust, ${Q\mathrm{map}}_{\nu}^{\mathrm{ds}}$ and ${U\mathrm{map}}_{\nu}^{\mathrm{ds}}$ are calculated by the following equation using thermal dust $Q$ and $U$ maps, ${Q\mathrm{map}}_{\nu}^{\mathrm{d}}$ and $ {U\mathrm{map}}_{\nu}^{\mathrm{d}}$ as 
\begin{eqnarray}
    {Q\mathrm{map}}_{\nu}^{\mathrm{ds}}
    &=& -\frac{B_{\nu}(T_{\mathrm{CMB}})}{B_{\nu}(T_{\mathrm{d}})} {Q\mathrm{map}}_{\nu}^{\mathrm{d}}, \nonumber\\
    {U\mathrm{map}}_{\nu}^{\mathrm{ds}}
    &=& -\frac{B_{\nu}(T_{\mathrm{CMB}})}{B_{\nu}(T_{\mathrm{d}})} {U\mathrm{map}}_{\nu}^{\mathrm{d}}. \label{shadow_map}
\end{eqnarray}
As discussed in Sect.\ref{sec:shadow}, we assume that the dust temperature is constant along each line-of-sight.

We take the thermodynamic temperature unit $\mathrm{K_{CMB}}$.
Then, the frequency dependence of the CMB signals disappears, and the CMB signals are able to be treated as a constant component in frequency space.
We apply a Gaussian smoothing with 2200' (FWHM) to the mock sky maps following \citet{Ichiki2019}.
We use the Galactic mask of the P06 mask \citep{Page_2007} degraded to $N_{\mathrm{side}}=4$ where the sky coverage is  $f_{\mathrm{sky}}=0.56$. 

\subsection{How to revise the Delta-map method} \label{subsec:delta_shadow}

We adopt the Delta-map method  \citep{Ichiki2019} for foreground removal.
The Stokes $Q$ at the frequency $\nu$ in the direction of $\hat{n}$ and the unit of thermodynamic temperature $\mathrm{K_{CMB}}$
is modeled as 
\begin{eqnarray}
    Q_{\nu}(\hat{n})
    &=& Q^{\mathrm{CMB}}(\hat{n}) 
    + g_{\nu} D^{\mathrm{d}}_{\nu}(\hat{n}) Q^{\mathrm{d}}_{\nu_*^{\mathrm{d}}}(\hat{n})
    + g_{\nu} D^{\mathrm{s}}_{\nu}(\hat{n}) Q^{\mathrm{s}}_{\nu_*^{\mathrm{s}}}(\hat{n}) \label{ObservedQ},
\end{eqnarray}
where $Q^{\mathrm{CMB}}(\hat{n})$, $Q^{\mathrm{d}}_{\nu_*^{\mathrm{d}}}(\hat{n})$ and $Q^{\mathrm{s}}_{\nu_*^{\mathrm{s}}}(\hat{n})$ are the Stokes $Q$ of the CMB,  thermal dust emission at the reference frequency $\nu_*^{\mathrm{d}}$, and synchrotron emission at the reference frequency $\nu_*^{\mathrm{s}}$ in the direction of $\hat{n}$, respectively. 
A factor $g_{\nu}$ is introduced to convert the unit from brightness to thermodynamic temperature. 
The $D^{\mathrm{d,s}}_{\nu}(\hat{n})$ is the frequency dependence of dust and synchrotron emission, respectively.
The original Delta-map method \citep{Ichiki2019} models the frequency dependence by the following equations:
\begin{eqnarray}
    D_{\nu}^{\mathrm{d}}(\beta_{\mathrm{d}}(\hat{n}), T_{\mathrm{d}}(\hat{n}))
    &=& \left( \frac{\nu}{\nu_*^{\mathrm{d}}} \right)^{\beta_{\mathrm{d}}(\hat{n})+1}
    \frac{e^{x_{\mathrm{d}*}(\hat{n})}-1}{e^{x_{\mathrm{d}}(\hat{n})}-1}, \label{dustfreq}\\
    D_{\nu}^{\mathrm{s}}(\beta_{\mathrm{s}}(\hat{n}))
    &=& \left( \frac{\nu}{\nu_*^{\mathrm{s}}} \right)^{\beta_{\mathrm{s}}(\hat{n})}, \label{synchfreq}
\end{eqnarray}
where $\beta_{\mathrm{d}}$ and $\beta_{\mathrm{s}}$ are spectral indices of dust and synchrotron emission, respectively,  $x_{\mathrm{d}}(\hat{n}) = h\nu / k_{\mathrm{B}} T_{\mathrm{d}}(\hat{n})$, and $x_{\mathrm{d}*}(\hat{n}) = h\nu_*^{\mathrm{d}} / k_{\mathrm{B}} T_{\mathrm{d}}(\hat{n})$.
These frequency dependencies do not include the effect of the CMB shadow.

In order to deal with the effect of the CMB shadow due to dust in the Delta-map method, the frequency dependence of dust $D^{\mathrm{d}}_{\nu}$ is modified to include 
not only the emission from the thermal dust but also the absorption of the CMB monopole due to dust 
as 
\begin{equation}
    D_{\nu}^{\mathrm{d}}(\beta_{\mathrm{d}}(\hat{n}), T_{\mathrm{d}}(\hat{n}))
    = \left( \frac{\nu}{\nu^{\mathrm{d}}_*} \right)^{\beta_{\mathrm{d}}(\hat{n})+1}
    (e^{x_{\mathrm{d}*}(\hat{n})}-1)
    \left(\frac{1}{e^{x_{\mathrm{d}}(\hat{n})}-1} - \frac{1}{e^{x}-1}\right), \label{dustfreqinshadow}
\end{equation}
where $x = h\nu / k_{\mathrm{B}}T_{\rm CMB}$. The essences of foreground removal with the Delta-map method are reviewed in Appendix \ref{sec:delta}.

\section{Results} \label{sec:results}
We apply the original Delta-map method \citep{Ichiki2019} 
and the revised Delta-map method introduced in Sect.\ref{subsec:delta_shadow} to the mock sky maps.

\begin{figure}[h]
  \begin{minipage}[b]{0.5\linewidth}
    \centering
    \includegraphics[keepaspectratio, scale=0.6]{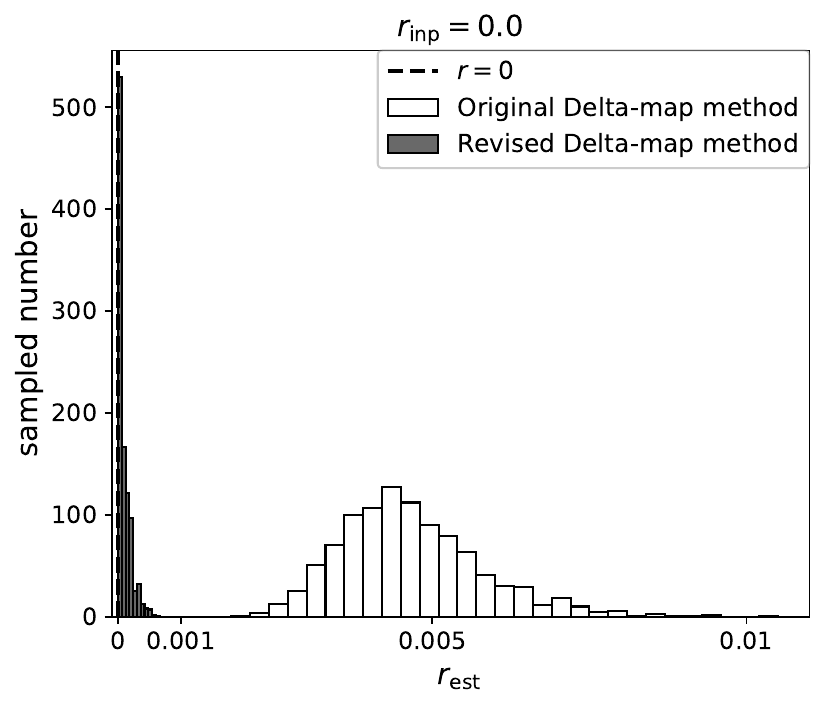}
  \end{minipage}
  \begin{minipage}[b]{0.5\linewidth}
    \centering
    \includegraphics[keepaspectratio, scale=0.6]{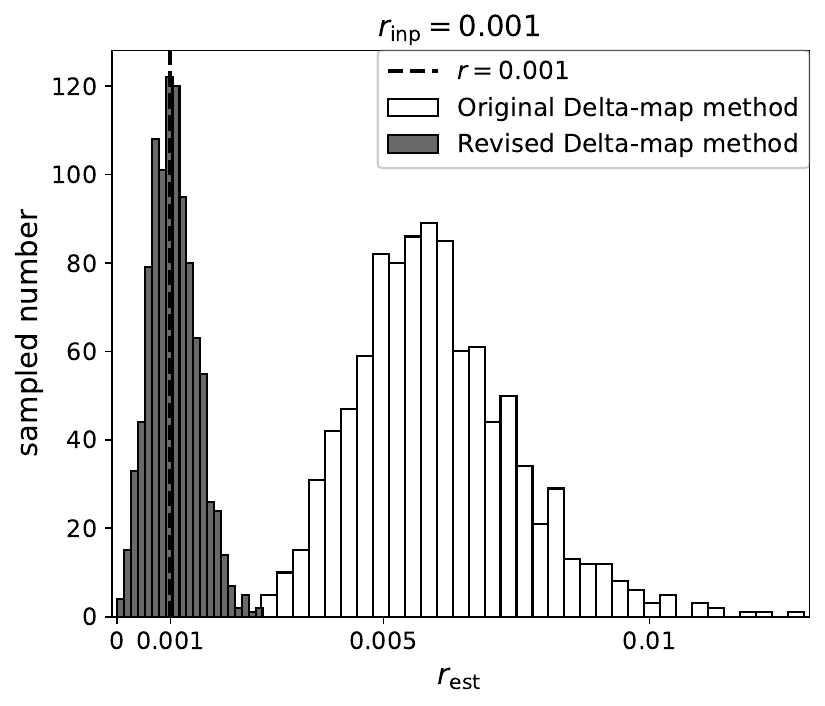}
  \end{minipage}
  \caption{
  The histograms of  $r_{\mathrm{est}}$ extracted from the 1000 realizations of the CMB maps when $r_{\mathrm{inp}}=0$ (left panel) and $r_{\mathrm{inp}}=0.001$ (right panel).
  The results from applying the original Delta-map method are shown by a white histogram.
  The results from applying the revised Delta-map method are shown by a gray histogram.
  }
    \label{fig:r_hist}
\end{figure}

In Fig.\ref{fig:r_hist}, we show the histograms of the estimated tensor-to-scalar ratio $r_{\mathrm{est}}$ from 1000 realizations of the CMB when $r_{\mathrm{inp}}=0$ (left) and $0.001$ (right).
When the input CMB maps do not include the CMB $B$-mode signals generated by primordial gravitational waves ($r_{\mathrm{inp}}=0$), the results from applying the original Delta-map method (white histogram) show that all obtained $r_{\mathrm{est}}$ have finite value ($\neq 0$), and we get $r_{\mathrm{est}}=(4.66\pm1.14)\times10^{-3}$ (68\%CL).
It indicates that if the CMB shadow is not taken into account properly in the foreground removal process, we are led to false detection of the CMB $B$-mode polarization signals generated by primordial gravitational waves.
The result from applying the revised Delta-map method is shown by a gray histogram.
About half of  $r_{\mathrm{est}}$ are distributed in the $r=0$ bin.
From the result, the upper limit on $r_{\mathrm{est}}$ is set to $r_{\mathrm{est}}<0.30\times10^{-3}$ (95\%CL).
It shows that the revised Delta-map method is able to avoid the false detection of the primordial gravitational wave origin CMB $B$-mode polarization signals.

When $r_{\mathrm{inp}}=0.001$,  $r_{\mathrm{est}}$ from applying the original Delta-map method (white histogram) is overestimated, and we get $r_{\mathrm{est}}=(5.99\pm1.53)\times10^{-3}$ (68\%CL).
It indicates that missing the CMB shadow in the foreground removal process leads to an overestimation of the amplitude of the tensor-mode perturbations. 
The result from applying the revised Delta-map method is shown by a gray histogram.
The obtained $r_{\mathrm{est}}$ concentrate around $r=0.001$, and we get $r_{\mathrm{est}}=(1.06\pm0.43)\times10^{-3}$ (68\%CL).
It shows that the revised Delta-map method is able to access $r=0.001$ by mitigating the CMB shadow. 

\begin{figure}[h]
    \centering
    \includegraphics[width=7cm,clip]{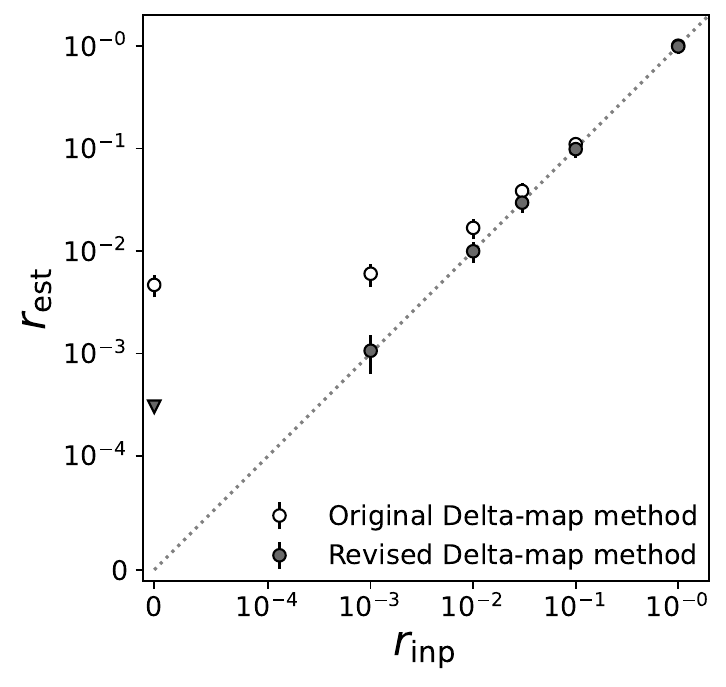}
    \caption{
    The mean values of $r_{\mathrm{est}}$ extracted from the original Delta-map method (open circles) and the revised Delta-map method (gray circles) for various $r_{\mathrm{inp}}$ in the range of $0.001\leq r_{\mathrm{inp}}\leq 1$.
    The error bars describe 68\%CL.
    In the range $r_{\mathrm{inp}}, r_{\mathrm{est}} < 10^{-4}$,  a linear scale is used.
    In the other ranges, we plot the horizontal and vertical axes with logarithmic scales.
    The gray inverted triangle shows the upper limit obtained by the results from the revised Delta-map method (95\%CL).
    }
    \label{fig:r_inp_est}
\end{figure}

$r_{\mathrm{est}}$ for various $r_{\mathrm{inp}}$ in the range of $0.001\leq r_{\mathrm{inp}}\leq 1$ are shown in Fig.\ref{fig:r_inp_est}.
We found that if the CMB shadow is not taken into account properly in the foreground removal method (see open circles in Fig.\ref{fig:r_inp_est}), $r_{\mathrm{est}}$ becomes systematically larger than $r_{\mathrm{inp}}$ for $r\lesssim 0.03$ which is the observational current upper limit of $r$ \citep{Tristram2022}.
We can see that the revised Delta-map method (see gray circles in Fig.\ref{fig:r_inp_est}) improves the contamination of the CMB shadow effect significantly.


\section{Discussion} \label{sec:conclusion}
We have estimated how the CMB shadow affects forthcoming high-precision CMB $B$-mode experiments realistically using noiseless, simulated sky maps for the first time.
The Delta-map method has been applied as a foreground removal method.
We show that  if the CMB shadow effects are not taken into account in the foreground removal process, future observations would lead  to the  false detection of  the CMB $B$-mode polarization signals originating from primordial gravitational waves.
The  revised Delta-map method is able to avoid the false detection of
the CMB $B$-mode polarization signals originating from primordial gravitational waves. 
We show that the effect of the CMB shadow can be mitigated to target $r=0.001$ by taking into account the CMB shadow properly in the Delta-map method.

In this paper, we used the reionization bump to extract $r$. However, the current upper limit on $r$ \citep{Tristram2022} is set from the observations targeted to the recombination bump. Since the amplitude of the CMB $B$-mode polarization imprinted by primordial gravitational waves around the recombination bump is more significant than that of the reionization bump, the accuracy degradation due to the CMB shadow for ground-based CMB observations  is less significant than the current studies \citep{Nashimoto2020}.  Therefore, the fact that the signature of the CMB shadow yet has not been detected yet, is consistent with our results. However, our results suggest that the CMB shadow becomes a non-negligible effect for forthcoming high-precision observations. 
Although only the Delta-map method was used in this study, we can conclude that all removal methods using the foreground spectral models will lead to false detection in the same way if the  CMB shadow is not properly taken into account. 

Most of the ground-based experiments target the recombination bump and cannot access the reionization bump.
Since the Delta-map method performs the foreground removal in pixels, not in the Fourier space, 
the signals with $r=0.001$ are extractable even from the partially covered maps without 
being contaminated by ambiguous $B$-mode \citep[e.g.,][]{Bunn_2003}. 
The method is potentially applicable to ground-based CMB polarization observations. 
Therefore, extending  our  revised Delta-map method to the  spatial resolution of $N_{\mathrm{side}}=64$ to access to the recombination bump is an interesting next task.

So far, we have assumed that the polarization fraction of the CMB shadow takes the same value as the polarization fraction of the thermal dust emission.  
This is only true when the dust temperature is constant along each line-of-sight. 
\citet{zelko2022} have reported the three-dimensional interstellar dust temperature distribution. 
They show a variation of the dust temperature along a line-of-sight. 
Then, the Stokes $Q$ and $U$ are not able to be expressed by Eqs.(\ref{RadTrans_pol_q}) and (\ref{RadTrans_pol_u}).
In this case, the polarization fraction of the thermal dust emission for Stokes $Q$ is written as
\begin{eqnarray}
    \Pi^{\mathrm{d},Q}_{\nu}
= \frac{Q_{\nu}^{\mathrm{d}}}{I_{\nu}^{\mathrm{d}}}
= \frac{\int ds (\alpha_{\nu}^x(s) - \alpha_{\nu}^y(s))\:B_{\nu}(T_{\mathrm{d}}(s))/2}{\int ds (\alpha_{\nu}^x(s) + \alpha_{\nu}^y(s))\:B_{\nu}(T_{\mathrm{d}}(s))/2},
\end{eqnarray}
where $I_{\nu}^{\mathrm{d}}$ and $Q_{\nu}^{\mathrm{d}}$ are the Stokes $I$ and $Q$ of the thermal dust emission.
On the other hand, the polarization fraction of the CMB shadow due to dust for Stokes $Q$ is given by
\begin{eqnarray}
    \Pi^{\mathrm{ds},Q}_{\nu}
= \frac{\int ds (\alpha_{\nu}^x(s) - \alpha_{\nu}^y(s))\:B_{\nu}(T_{\mathrm{CMB}})/2}{\int ds (\alpha_{\nu}^x(s) + \alpha_{\nu}^y(s))\:B_{\nu}(T_{\mathrm{CMB}})/2}
=\frac{\int ds (\alpha_{\nu}^x(s) - \alpha_{\nu}^y(s))}{\int ds (\alpha_{\nu}^x(s) + \alpha_{\nu}^y(s))}.
\end{eqnarray}
The polarization fraction of the thermal dust emission and that of the CMB shadow due to dust do not take the same value in general. 
In such an analysis, spatial variation of the polarization direction of the  thermal dust emission in a line-of-sight  must be taken into account  \citep{Tassis+15} at the same time.
Foreground removal methods that account for the spatial variation of the dust temperature and polarization fraction of thermal dust emission are desired \citep[e.g.,][]{cMILC2021,Vacher+22}. 


\begin{acknowledgments}
This work is supported by MEXT KAKENHI Grant Number JP18H05539, JP20KK0065, JP22H04913, KAKENHI JP21K03585, and also supported  by JSPS Core-to-Core Program JPJSCCA20200003.
The work of TM was supported by the Graduate Program on Physics for the Universe (GP-PU), Tohoku University.
MN acknowledges support from JSPS KAKENHI Grant Number JP22J00388.
The work of KI was supported by KAKENHI JP18K03616, JP21H04467, and FOREST Program JPMJFR20352935.
We are greatly indebted to Kenji Amazaki for valuable discussion on dust stimulated emission.
The anonymous referee's comments were very valuable for brushing up the contents of the paper.
\end{acknowledgments}

%






\appendix
\section{Delta-map method} \label{sec:delta}
We review the foreground removal process of the Delta-map method in this section.
In the Delta-map method, we take into account sky-average and the first-order spatial fluctuations of foreground parameters: spectral indices of dust $\beta_{\rm d}$ and synchrotron emission $\beta_{\rm s}$, and the dust temperature $T_{\rm d}$.
Then, Equation (\ref{synchfreq}) are reduced to
\begin{eqnarray}
    D_{\nu}^{\mathrm{s}}(\beta_{\mathrm{s}}(\hat{n}))
    &\sim& D_{\nu}^{\mathrm{s}}(\bar{\beta}_{\mathrm{s}})
    + \frac{\partial D_{\nu}^{\mathrm{s}}}{\partial \beta_{\mathrm{s}}} \delta{\beta_{\mathrm{s}}(\hat{n})}
\end{eqnarray}
and Eqs.(\ref{dustfreq}) and (\ref{dustfreqinshadow}) are reduced to
\begin{eqnarray}
        D_{\nu}^{\mathrm{d}}(\beta_{\mathrm{d}}(\hat{n}), T_{\mathrm{d}}(\hat{n}))
    &\sim& D_{\nu}^{\mathrm{d}}(\bar{\beta}_{\mathrm{d}}, \bar{T}_{\mathrm{d}})
    + \frac{\partial D_{\nu}^{\mathrm{d}}}{\partial \beta_{\mathrm{d}}} \delta \beta_{\mathrm{d}}(\hat{n})
    + \frac{\partial D_{\nu}^{\mathrm{d}}}{\partial T_{\mathrm{d}}} \delta T_{\mathrm{d}}(\hat{n}),
\end{eqnarray}
where $\bar{\beta}_{\mathrm{s}}$, $\bar{\beta}_{\mathrm{d}}$, and $ \bar{T}_{\mathrm{d}}$ describe sky-average of each parameter, and $\delta \beta_{\mathrm{s}}(\hat{n})$, $\delta \beta_{\mathrm{d}}(\hat{n})$, and $\delta T_{\mathrm{d}}(\hat{n})$ describe the first-order variation.

The cleaned map is given by the following equation:
\begin{eqnarray}
    \vec{m}
    &=& \frac{[{Q\mathrm{map}}_{\nu_{\mathrm{CMB}}}, {U\mathrm{map}}_{\nu_{\mathrm{CMB}}}] 
    + \Sigma_{j=1}^5 \lambda_{\nu_j} [{Q\mathrm{map}}_{\nu_j}, {U\mathrm{map}}_{\nu_j}]}
    {1 + \Sigma_{j=1}^5 \lambda_{\nu_j}}, \label{Cleaned}
\end{eqnarray}
where $\nu_{\mathrm{CMB}}=140$ GHz is used as the frequency of the CMB channel, $\nu_{j}$ are the frequency bands of the foreground channels, which are 40, 60, 230, 280, and 340 GHz in this paper, 
and $\lambda_{\nu_j}$ are coefficients like the Lagrange multiplier coefficients.
The $Q\mathrm{map}_{\nu}$ and $U\mathrm{map}_{\nu}$ are  maps of the Stokes $Q$ and $U$ prepared in Sect.\ref{subsec:setup}.  
In order to remove foreground components, $\lambda_{\nu_j}$ is set, so that foreground terms in Equation (\ref{Cleaned}) vanish, that is $\vec{d}_{\nu_{\mathrm{CMB}}} + \mathbf{A} \vec{\lambda} = 0$, where
\begin{eqnarray}
    \vec{\lambda}
    &=& (\lambda_{\nu_1}, \lambda_{\nu_2}, ..., \lambda_{\nu_5})^T, \\
    \vec{d_{\nu}}
    &=& \left(D_{\nu}^{\mathrm{s}}(\bar{\beta}_{\mathrm{s}}), 
    \frac{\partial D_{\nu}^{\mathrm{s}}}{\partial \beta_{\mathrm{s}}}, 
    D_{\nu}^{\mathrm{d}}(\bar{\beta}_{\mathrm{d}}, \bar{T}_{\mathrm{d}}), 
    \frac{\partial D_{\nu}^{\mathrm{d}}}{\partial \beta_{\mathrm{d}}}, 
    \frac{\partial D_{\nu}^{\mathrm{d}}}{\partial T_{\mathrm{d}}}\right)^T \\
    \mathbf{A}
    &=& (\vec{d}_{\nu_1}, \vec{d}_{\nu_2}, ..., \vec{d}_{\nu_5}).
\end{eqnarray}
So, $\lambda_{\nu_j}$ is obtained by $\vec{\lambda} = -\mathbf{A}^{-1}\vec{d}_{\nu_{\mathrm{CMB}}}$.

For the process of estimating parameters: tensor-to-scalar ratio $r$, the spatial average of foreground parameters $\bar{\beta}_{\mathrm{s}}, \bar{\beta}_{\mathrm{d}}$, and  $\bar{T}_{\mathrm{d}}$, we follow \citet{Ichiki2019}, who performs
maximum likelihood estimation using 
the log-likelihood  given by
\begin{eqnarray}
    \ln\mathcal{L}
    &=& -\frac{1}{2} \left(\ln |2\pi \mathbf{C}| + \vec{m}^T \mathbf{C} \vec{m}\right).
\end{eqnarray}
We use the same covariance matrix $\mathbf{C}$ as \citet{Ichiki2019}.

\section{Stimulated emission of dust grains} \label{sec:stiemi}


Some astrophysicists \citep[e.g.,][]{Harwit} claimed that the net absorption coefficient of the ambient radiation fields by dust grains should be
attenuated due to the stimulated emission by a factor of $[1-\exp(-h\nu/k_\mathrm{B}T_\mathrm{d})]$ when dust grains are in local thermal equilibrium at a temperature $T_\mathrm{d}$. 
In the microwave region, $[1-\exp(-h\nu/k_\mathrm{B}T_\mathrm{d})]\sim h\nu/k_\mathrm{B}T_\mathrm{d}$ is much smaller than 1. The attenuation of the net absorption coefficient is prominent.
Since the optical properties of dust grains are well 
described by the dielectric media model, 
the temperature dependence of the absorption coefficient of the dielectric media provides experimental evidence that the reduction of the absorption coefficient of dust grains due to the stimulated emission does not happen. 
\citet{Otsuka:21} measured the temperature dependence of the absorption coefficient in millimeter wave bands for various 
dielectric media with the Fourier transform spectrometer. 
They used blackbody radiation with 6000 K as a light source 
and measured the transmittance in millimeter wave bands. 
The measurements were performed for sample temperatures of 77 K and 300 K. 
If the stimulated emission works as discussed above,  
the absorption coefficient of the dielectric media at 77 K
must be a factor of 3 larger than those at 300 K.
The results reported by \citet{Otsuka:21} were contrary to the expectation. 
For all the measured samples, the absorption coefficients at 77K did not exceed those at 300K. 
This indicates that the attenuation of the absorption coefficient in the microwave region due to the stimulated emission does not happen in dielectric media that are dust grains.  

Physical considerations of why the reduction factor due to the stimulated emission does not have to be taken into account in the absorption coefficient of dust grains have been made by \citet{Bond} and \citet{Wright}. 
The main point of the issue is that dust grains absorb photons by scattering or creating phonons that are bosonic quasiparticles in optically active vibration bands. 
The textbook model of the two-level system of the absorbers adapted to 
evaluate the effect of the stimulated emission \citep{radipro} is not appropriate for dust grains.



\bibliography{0_ref}{}
\bibliographystyle{aasjournal}



\end{document}